\documentclass[reprint,twocolumn,notitlepage,prb,superscriptaddress,longbibliography,amsmath,amssymb,floatfix]{revtex4-2}

\usepackage{graphicx}
\usepackage{amsfonts}
\usepackage{color}
\usepackage[colorlinks=true,linkcolor=blue,citecolor=blue]{hyperref}
\usepackage{soul}
\usepackage{physics}
\usepackage{chemformula}

\begin{document}

\title{Quantum supercritical crossover with dynamical singularity}

\author{Junsen Wang}
\affiliation{Institute of Theoretical Physics, Chinese Academy of Sciences, Beijing 100190, China}
\affiliation{Anhui Province Key Laboratory of Condensed Matter Physics at Extreme Conditions, High Magnetic Field Laboratory, Chinese Academy of Sciences, Hefei 230031, China}
\author{Enze Lv}
\affiliation{Institute of Theoretical Physics, Chinese Academy of Sciences, Beijing 100190, China}
\affiliation{School of Physical Sciences, University of Chinese Academy of Sciences, Beijing 100049, China}
\author{Xinyang Li}
\affiliation{Institute of Theoretical Physics, Chinese Academy of Sciences, Beijing 100190, China}
\affiliation{Peng Huanwu Collaborative Center for Research and Education, Beihang University, Beijing 100191, China}
\author{Yuliang Jin}
\email{yuliangjin@itp.ac.cn}
\affiliation{Institute of Theoretical Physics, Chinese Academy of Sciences, Beijing 100190, China}
\affiliation{School of Physical Sciences, University of Chinese Academy of Sciences, Beijing 100049, China}
\author{Wei Li}
\email{w.li@itp.ac.cn}
\affiliation{Institute of Theoretical Physics, Chinese Academy of Sciences, Beijing 100190, China}
\affiliation{School of Physical Sciences, University of Chinese Academy of Sciences, Beijing 100049, China}

\date{\today}

\begin{abstract}
Bounded by crossover lines exhibiting universal scaling, the supercritical regime above the critical endpoint is characterized by strong fluctuations and intriguing phenomena. In this study, we extend this notable concept of supercritical crossover to the quantum critical endpoint (QCEP), by studying the prototypical mixed-field quantum Ising and Potts models through tensor network calculations and scaling analyses. We reveal the existence of quantum supercritical (QSC) crossover lines, determined by not only response functions but also quantum information quantities, near the QCEP. A supercritical scaling law, $h \sim (g - g_c)^{\Delta}$, is found, where $g$ ($h$) is the transverse (longitudinal) field, $g_c$ is the critical field, and $\Delta$ is the so-called gap exponent of the QCEP. Moreover, we demonstrate that the QSC crossover line acts as a boundary for the emergence of dynamical singularities in quench dynamics. This singularity manifests as a distinctive cusp with a critical exponent of 1/2, signaling a new dynamical universality class. We also propose utilizing Rydberg atom arrays as an experimental platform to observe these QSC crossovers and dynamical singularities. Our work establishes a theoretical framework for understanding the role of QCEP and associated supercritical crossovers in both equilibrium and non-equilibrium quantum many-body systems.
\end{abstract}

\maketitle

\section{Introduction}
Above the critical endpoint (CEP) situated at the terminus of a first-order transition line, there arise classical supercritical fluids~\cite{Cagniard1822, Andrews1869}.
In the supercritical regime, distinct phase transitions are replaced by continuous crossovers with gradual property changes. The supercritical crossover boundaries have been extensively studied in the context of liquid-gas transitions, from aspects such as thermodynamic response functions, classical dynamics, symmetry and critical universality~\cite{fisher1969, xu2005, ruppeiner2012, may2012, brazhkin2012, brazhkin2013, luo2014, gallo2014, li2024}. Beyond their original discovery in classical fluids, CEPs and their associated thermal supercritical phenomena are now known to be ubiquitous, as extensively studied in magnetic materials~\cite{wang2023}, Mott insulators and superconductors~\cite{sordi2012, sordi2013, vucicevic2013, downey2023}, and quark-gluon plasma in high-energy physics~\cite{sordi2024}.

On the other hand, while the quantum critical endpoint (QCEP), which terminates a first-order quantum phase transition line, has recently been experimentally observed, its related effects remain an open area of study. For example, in the itinerant metamagnet \ch{Sr_3Ru_2O_7}, its CEP can be depressed toward zero temperature, i.e., QCEP, by a magnetic field~\cite{grigera2001}. Similar situation also happens for metamagnetic transitions in heavy fermion compound \ch{UTe_2}~\cite{wu2025}, ferromagnetic superconductor \ch{UGe_2}~\cite{taufour2010} and antiferromagnetic insulator \ch{Na_2Co_2TeO_6}~\cite{arneth2024}. Moreover, nowadays quantum simulators also become an ideal platform for study quantum Ising systems~\cite{schauss2015, Guardado-Sanchez2018, keesling2019, scholl2021, ebadi2021}, where QCEP can emerge through parameter fine-tuning. Inspired by the rich thermal supercritical phenomena above the CEP, it is natural to suspect the existence of quantum supercritical (QSC) states near the QCEP and related crossover phenomena driven by quantum rather than thermal fluctuations --- a central topic of this work.

\begin{figure}[t] 
\centering
\includegraphics[width=0.46\textwidth]{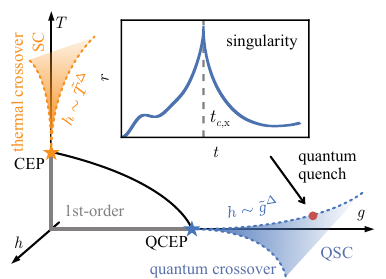} 
\caption{
\textbf{Critical endpoint and supercritical regime.}
In the $h$-$T$ plane, there exist supercritical (SC) states above the thermal critical endpoint (CEP), enclosed by the crossover lines $h\sim \tilde{T}^\Delta$~\cite{li2024}, with $\tilde{T} \equiv T-T_c$ and $\Delta\equiv\beta+\gamma$ the gap exponent. In the zero-temperature $g$-$h$ plane, there exists quantum critical endpoint (QCEP) and quantum supercritical (QSC) crossover lines with scaling law $h \sim \tilde{g}^{\Delta}$, where $\tilde{g} \equiv g-g_c$. The inset reveals a dynamical singularity when the system Hamiltonian is quenched to the QSC crossover line, and the Loschmidt rate function $r$ (see definition in the main text) exhibits a singular 1/2 cusp at the critical time $t_{c, \rm{x}}$.
}
\label{fig1}
\end{figure}

Below, we address this question from both equilibrium and non-equilibrium aspects, through comprehensive studies of the paradigmatic and minimal quantum Ising and Potts model. Our main conclusions are summarized in Fig.~\ref{fig1}, where we find in the zero-temperature $g$-$h$ plane a quantum analogue of the classical supercriticality in the $h$-$T$ plane. By examining response functions and quantum information quantities, we unveil the quantum supercritical (QSC) crossover lines that adhere to the scaling law $h \sim (g-g_c)^{\Delta}$, determined by the universality class of QCEP. Notably, we discover that these crossovers, although corresponding to smooth changes in equilibrium properties, can give rise to singularities in Loschmidt rate function of quantum quench dynamics (see inset of Fig.~\ref{fig1}). Intriguingly, we find a cusp singularity with exponent of 1/2 rather than the conventional 1 (i.e., linear cusp), giving rise to a new universality class of dynamical quantum phase transition (DQPT)~\cite{heyl2013}. Prior studies have shown that  DQPTs can decouple from equilibrium quantum phase transitions~\cite{vajna2014}. In contrast, the 1/2-cusp DQPTs revealed in this work are intimately connected to the universal QSC crossover lines emanating from the QCEP.  Specifically speaking, one can regard the QSC crossover lines as the ``dynamically singular phase boundaries'', in sharp contrast to classical supercritical crossover where dynamical singularity is absent. Finally, we discuss experimental platforms for investigating QSC crossovers and related phenomena.

\section{Equilibrium characterization of QSC crossover}
As a prominent example, we consider the one-dimensional (1D) mixed-field quantum Ising model (MFQIM), described by the following Hamiltonian,
\begin{equation}
    \label{tficwlf}
    H(g, h) = - J \sum_{\langle i,j\rangle} S_i^z S_j^z  - g\sum_i S_i^x - h \sum_i S_i^z,
\end{equation}
where $S_i^\mu$ represents the spin-1/2 operator at site $i$, along direction $\mu=x,z$. $J=1$ is the FM exchange between nearest-neighboring sites $\langle i,j\rangle$, set as the energy scale. The transverse field $g$ introduces quantum fluctuations in the system and induces a QCEP at $g_c=1$ (i.e., $\tilde{g} \equiv g - g_c = 0$). This QCEP marks the termination of a first-order line induced by the longitudinal field $h$, as depicted in the $g$-$h$ quantum phase diagram of Fig.~\ref{fig1}.

To study equilibrium properties associated with the QSC crossover, we employ the variational uniform matrix product state algorithm~\cite{zaunerStauber2018} to simulate the ground-state properties of the 1D MFQIM directly in the thermodynamical limit. In the appendix, we also give numerical results of 2D infinite cylinders with width $W=12$ of the MFQIM and the 1D $q=3,4$-state quantum Potts models, all belonging to different universality classes.

\begin{figure*}[t]
\centering
\includegraphics[width=.98\textwidth]{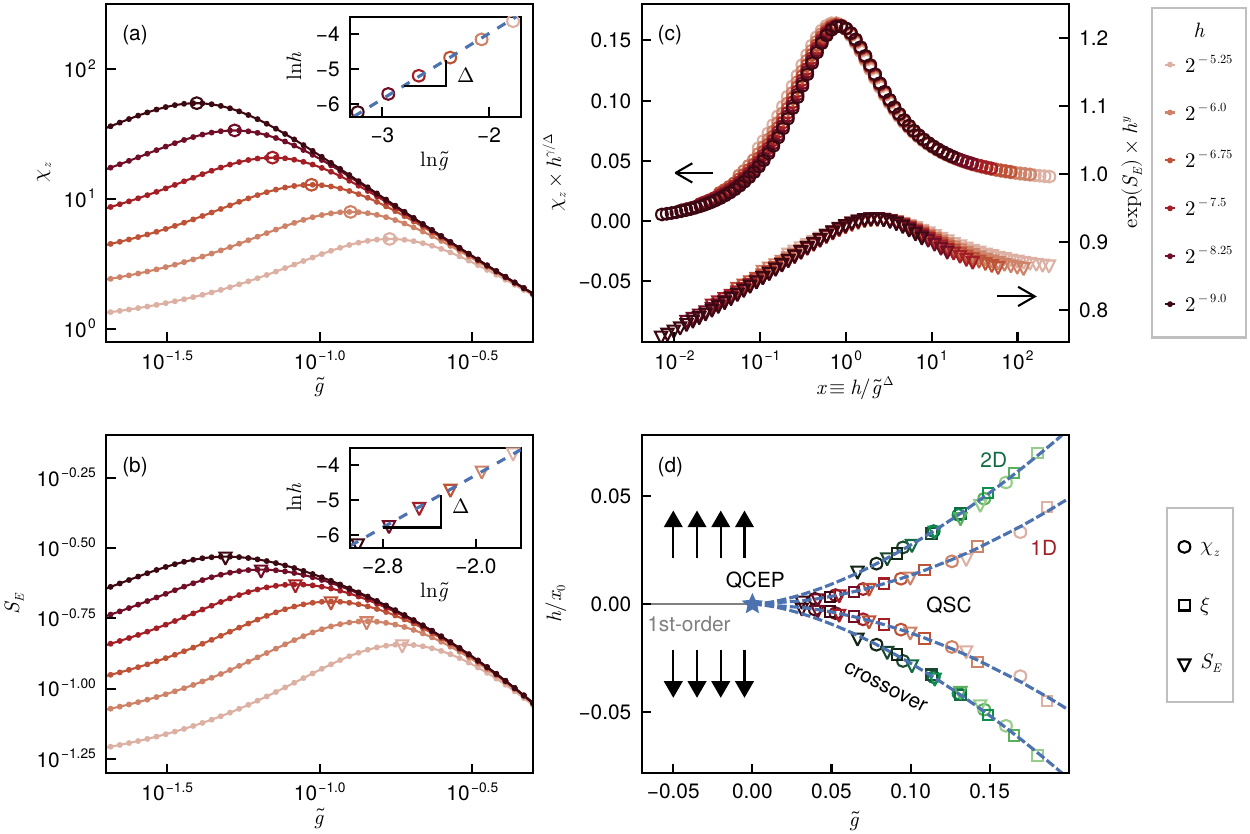}
\caption{\textbf{Determination of the QSC crossover in equilibrium and the supercritical quantum phase diagram.} magnetic susceptibility $\chi_z$ (a) and entanglement entropy $S_E$ (b) used to determine the QSC crossover. In the insets, the hollow symbols represent the corresponding peaks, demonstrating a power-law scaling $h \sim \tilde{g}^{\Delta}$, as indicated by the blue dashed line, with the gap exponent $\Delta = \beta + \gamma = 1.875$. (c) Data collapse for these quantities. The peaks collapse to a single point at $x = x_0$, which is the maximum of the corresponding universal scaling function. Note that for $S_E$, we find the best collapse is obtained by setting $y = 0.058$, slightly greater than $c\nu/6\Delta \simeq 0.044$, which is possibly due to nonnegligible contribution from the regular part. (d) The QCEP and associated QSC states. A first-order line separates the spin-up and spin-down states, and the spin up-down indistinguishable QSC states are enclosed by two crossover lines $h/x_0 = \tilde{g}^\Delta$ determined from various quantities. Universal crossover lines of both 1D and 2D results are given, with different gap exponent.}
\label{fig2}
\end{figure*}

In Fig.~\ref{fig2}(a), we present the magnetic susceptibility $\chi_z \equiv \partial m_z/\partial h$ as a function of $\tilde{g}$ under various small fixed longitudinal fields $h$. All curves initially exhibit an algebraic divergence following $\chi_z \sim \tilde{g}^{-\gamma}$ with $\gamma= 7/4$. However, $\chi_z$ eventually deviates from universal scaling behaviors, reaching a maximum before getting suppressed. Moreover, these maxima shift towards greater $g$ as $h$ increases, marked as hollow circles, which define a crossover line. The inset demonstrates that this line follows a power-law scaling $h \sim \tilde{g}^{\Delta}$ with $\Delta \equiv \beta+\gamma = 15/8$, dubbed the \emph{QSC scaling law}. Similarly, in Fig.~\ref{fig2}(b), we compute the bipartite entanglement entropy $S_E(g)$ under various fixed $h$. Each initially exhibits a logarithmic divergence, $S_E \sim -\frac{c\nu}{6}\ln\tilde{g}$, with central charge $c=1$ and correlation length exponent $\nu=1$. This is followed by a peak before eventual suppression. By collecting the peak positions, we determine a crossover line that adheres to the QSC scaling law, as shown in the inset of Fig.~\ref{fig2}(b).

The QSC scaling law can be understood through a scaling analysis~\cite{carr2011,sachdev2015,continentino2017}. In the vicinity of a QCEP, we suppose that the magnetic susceptibility is a generalized homogeneous function $\chi_z \sim  h^{-{\gamma}/{\Delta}} \phi_{\chi_z}(x)$, where $x\equiv h/\tilde{g}^{\Delta}$. Although the scaling form is derived from renormalization group analysis~\cite{fisher1974, pelissetto2002, hankey1972, kirkpatrick2015}, the specific form of universal scaling function $\phi_{\chi_z}(x)$ needs to be determined numerically. The data collapse results are shown in Fig.~\ref{fig2}(c). There is a single peak at $x=x_0$, with ${\phi}'_{\chi_z}(x_0)=0$, which is right at the QSC crossover line identified in Fig.~\ref{fig2}(a). Therefore, the universal scaling law $h / \tilde{g}^{\Delta} = x_0$ is inherently encoded in the scaling function $\phi_{\chi_z}(x)$. Noteworthily, the QSC crossovers are not determined by changes in short-range correlations (regular part), as in conventional crossovers, but rather by universal behaviors near QCEP (singular part). For entanglement entropy, it scales as $S_E \sim \frac{c}{6}\ln \xi$ near the 1D QCEP according to Calabrese and Cardy~\cite{calabrese2004}. Since the singular part of correlation length $\xi$ close to the QCEP can be described by a generalized homogeneous function~\cite{fisher1974,pelissetto2002,hankey1972,kirkpatrick2015}, $\xi \sim \tilde{g}^{-\nu}\phi_{\xi}(x)$, where $x\equiv h/\tilde{g}^{\Delta}$, we then have that $S_E \sim - \frac{c \nu}{6} \ln \tilde{g} \sim - \frac{c \nu}{6 \Delta} \ln h$ naturally leads to a conjectured scaling form $\exp(S_E) \sim h^{-c\nu/6\Delta}\phi_{S_E}(x)$ near the 1D QCEP~\cite{vanhecke2019}, with $\phi_{S_E}(x)$ a universal scaling function. In Fig.~\ref{fig2}(c), data collapse gives this scaling function, with the QSC crossovers located at its peak.

In Fig.~\ref{fig2}(d), we summarize the supercritical quantum phase digram derived from various physical quantities. The one corresponding to the correlation length and all 2D results are given in appendix. The determined crossovers have the same form $h/x_0 = \tilde{g}^{\Delta}$, separating the spin-up (``liquid"), spin-down (``gas"), and the up-down indistinguishable (``supercritical fluid") regimes. Note that the 1D and 2D QSC crossover lines are distinct due to their different universality classes.

\section{Dynamical singularity of QSC crossover}
In the study of supercritical fluids, particle dynamics constitute an important aspect and have been used to determine the dynamic crossovers~\cite{xu2005,brazhkin2012,brazhkin2013}. Here we consider quantum quench process, and reveal the emergence of dynamical singularity for QSC crossovers. Specifically, the system is initialized in the ground state $|\psi_i\rangle$ of a pre-quenched Hamiltonian {$H_i = H(g_i, h)$}, and its subsequent time evolution is driven by a post-quenched Hamiltonian {$H_f = H(g_f, h)$}. The central quantity is the Loschmidt rate function (LRF), 
\begin{equation}
    r(t) = - \lim_{L\rightarrow \infty} \frac{1}{L} \ln|G(t)|^2,
\end{equation}
where $L$ is the system size, and $G(t) = \langle \psi_i | e^{-iH_ft} | \psi_i \rangle$ is the Loschmidt amplitude~\cite{quan2006,silva2008}. According to the seminal work of Heyl \emph{et al.}~\cite{heyl2013}, the formal similarity between the canonical partition function and the Loschmidt amplitude motivates the definition of dynamical quantum phase transition (DQPT) as the occurrence of non-analytical behavior in LRF. In the following, we utilize this ``dynamical free energy'', defined for real time, to investigate 
the quantum quench dynamics near the QSC crossovers.

\begin{figure*}
\centering
\includegraphics[width=.98\textwidth]{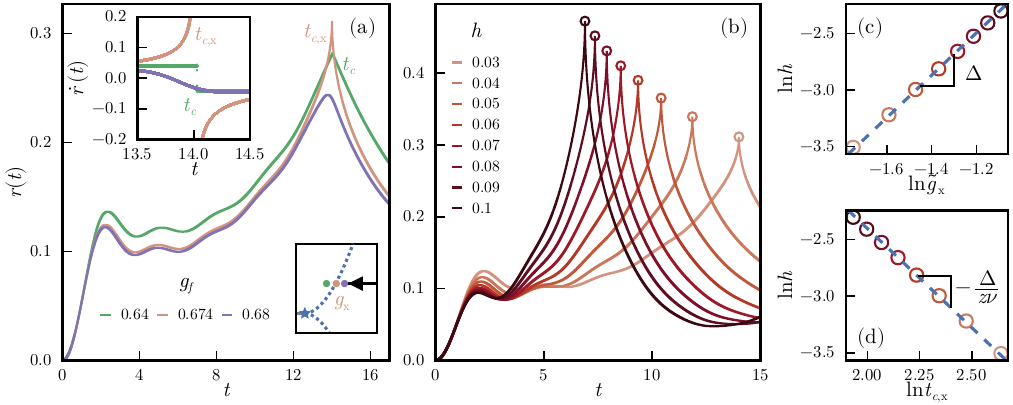}
\caption{\textbf{Dynamical singularity of QSC crossover in the 1D MFQIM.} (a) LRF $r(t)$ for quantum quench from $g_i=\infty$ to various $g_f$ with fixed $h=0.03$. The lower inset illustrates the quench protocol, where the asterisk marks the QCEP, and the dashed line indicates the QSC crossover. The upper inset shows its time derivatives $\dot{r}$, where linear cusp at $t_c$ and singular cusp at $t_{c,\text{x}}$ can be identified. (b) LRF with a singular cusp at $t_{c,\text{x}}$ for various $h$ and $g_f = g_{\text{x}}(h)$. (c) Log-log plot of $h$ vs. $\tilde{g}_{\text{x}}$, showing the QSC scaling law $h \sim \tilde{g}_{\text{x}}^\Delta$. (d) Log-log plot of $h$ vs. $t_{c,\text{x}}$, illustrating a scaling relation $h \sim (1/t_{c,\text{x}})^{\Delta/z\nu}$, where $\Delta$ and $z\nu$ are critical exponents of (1+1)D Ising universality class.}
\label{fig4}
\end{figure*}

In Fig.~\ref{fig4}(a), the time evolution block decimation algorithm~\cite{vidal2007,orus2008} is used to calculate LRF directly in the thermodynamic limit. (More details on this method is given in appendix.) We consider a quantum quench from $g_i=\infty$, i.e., $|\psi_i \rangle = | \rightarrow, \rightarrow, ... \rightarrow\rangle$, to various final $g_f$ values, with fixed $h=0.03$ [lower inset of Fig.~\ref{fig4}(a)]. For large $g_f$, the LRF displays rounded peaks in the time domain considered; while for small \( g_f \), the LRF develops a cusp at a critical time $t_c$, accompanied by a discontinuous jump in its time derivative $\dot{r}$ [upper inset of Fig.~\ref{fig4}(a)]. The transition from a rounded peak to a cusp occurs at a special value $g_f = g_{\text{x}}$. When quenching precisely to it, a singular cusp with critical exponent $\alpha = 1/2$ (see next section for more details) emerges
at a critical time $t_{c, \text{x}}$. Correspondingly, the time derivative $\dot{r}(t)$ exhibits a \emph{divergent} discontinuity. At the critical time $t_x$ or $t_{c, \text{x}}$, the sharp cusp in LRF illustrates vividly the distinction between different regimes separated by the QSC crossover line from a dynamical perspective. For $h=0$, the linear cusp can be explained as the unstable fixed point of the equilibrium Ising model~\cite{heyl2015}. Since $h$ is a relevant perturbation, it should change the universality class and thus account for the distinct Loschmidt-rate exponent of 1/2. Nevertheless, an explicit renormalization group analysis is needed to comprehend the numerical results in the future.

Since $g_{\text{x}}$ is the boundary that separates regime with and without dynamical singularity, we systematically determine it for different small $h$. The results are presented in Fig.~\ref{fig4}(b), and all of them show a singular 1/2-cusp. Remarkably, this boundary follows the QSC scaling law $h \sim \tilde{g}_{\text{x}}^{\Delta} $, where $\tilde{g}_{\text{x}} = g_{\text{x}} - g_c$, as illustrated in Fig.~\ref{fig4}(c), indicating that $g_{\text{x}}$ lies precisely on a QSC crossover line and $\Delta$ is the critical exponent of (1+1)D Ising universality class. DQPT was initially discovered in quenches across $g_c$~\cite{heyl2013}, and instances of DQPT independent of equilibrium quantum phase transitions have also been reported~\cite{vajna2014, canovi2014, andraschko2014, markus2015, zunkovic2018}. Here we find in quantum Ising model the DQPT singularities remain governed by the universality class of QCEP. Unlike the linear cusp observed in quenches across QCEP~\cite{heyl2013}, this singular cusp arises at QSC crossovers due to the relevant perturbation $h$, establishing a distinct dynamical universality class~\cite{heyl2015}. Moreover, the first critical time $t_{c, \text{x}}$ for these singular cusps also obeys a scaling relation, $h \sim (1/t_{c, \text{x}})^{\Delta/z\nu}$, as demonstrated in Fig.~\ref{fig4}(d). Notably, based on our scaling relation such singular cusp cannot be observed for $h=0$, as it appears at infinitely long time when quenching precisely to the QCEP, which is consistent with previous exact results~\cite{heyl2013}.

\begin{figure} 
\centering
\includegraphics[width=0.46\textwidth]{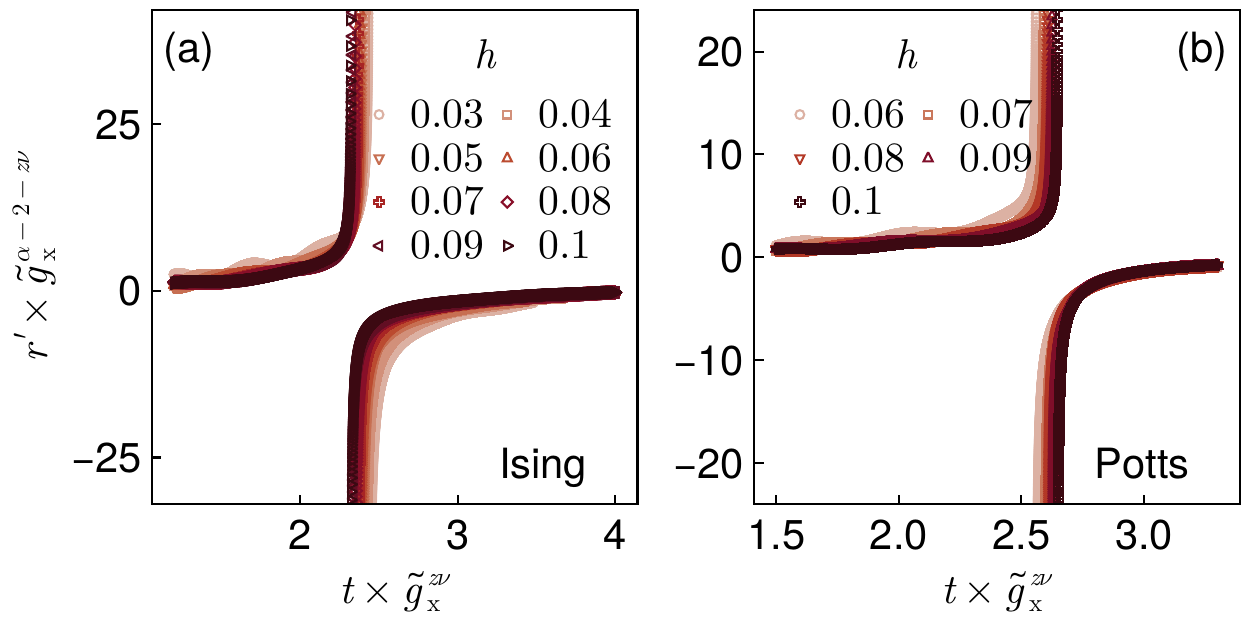} 
\caption{\textbf{Universal function of Loschmidt rate function near DQPT.} 
Data collapse for the time derivative of Loschmidt rate function near the first critical time $t=t_{c,\text{x}}$ for (a) 1D MFQIM and (b) 3-state MFQPM.}
\label{fig:sm_fig4} 
\end{figure}

\section{Universal scaling form of Loschmidt rate function near $t_{c,\rm{x}}$}\label{seclrf}
Here we give an intuitive understanding for the universal scaling relations given in Fig.~\ref{fig4} for the Ising model which is also verified in appendix for the Potts model, namely $h\sim \tilde{g}_{\text{x}}^\Delta$ and $h \sim (1/t_{c,\text{x}})^{\Delta/z\nu}$. According to the renormalization group theory, it is well-known that near a QCEP, the singular part of the free energy density takes the form $f = |\tilde{g}|^{2-\alpha} \phi_f\left(\frac{T}{|\tilde{g}|^{\nu z}}, \frac{h}{|\tilde{g}|^\Delta}\right)$~\cite{continentino2017}. Since the Loschmidt rate function $r(t)$ can be viewed as a dynamical analogue of the free energy, it is natural to expect that such a scaling form also holds near a dynamical singularity. Explicitly speaking, we conjecture that near $t=t_{c,\text{x}}$, the singular part of the Loschmidt rate function takes the form
\begin{equation}
	r_s = |\tilde{g}|^{2-\alpha} \phi_r\left(t|\tilde{g}|^{\nu z}, \frac{h}{|\tilde{g}|^\Delta}\right).\label{rs}
\end{equation}
Above equation directly leads to the two scaling relations found numerically, since the first dynamical singularity at $t = t_{c,\text{x}}$ corresponds to the position where both arguments in Eq.~\eqref{rs} remain as constants. To verify this scaling form, we first take time derivative of Eq.~\eqref{rs} to eliminate the regular contribution to the Loschmidt rate function, which leads to
\begin{equation}
	r'(t) = \tilde{g}^{2-\alpha+\nu z} \tilde{\phi}_r\left(t\tilde{g}^{\nu z}, \frac{h}{\tilde{g}^\Delta}\right),
\end{equation}
where $\tilde{\phi}_r = \partial_x \phi_r(x,y)$. We then perform data collapse for this quantity near $t=t_{c,\text{x}}$, as shown in Fig.~\ref{fig:sm_fig4} for both the 1D MFQIM and 1D 3-state MFQPM. It indeed confirms the expected universal scaling behavior.

\section{QSC crossover and Lee-Yang zeros}
\begin{figure}
	\centering
	\includegraphics[width=0.46\textwidth]{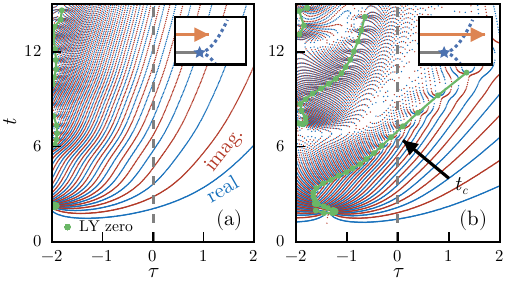}
	\caption{\textbf{Lee-Yang zeros of dynamical singularity for the 1D MFQIM.} Zeros of real (imaginary) part of boundary partition function $Z(z)$ are shown as blue (red) dots for (a) $g_f=0.4$ and (b) $g_f=0.65$ with $h=0.01$, $g_i=0$ and chain length $L=32$. The green dots indicate the LY zeros where the real and imaginary zero lines cross. The green lines are guide to the eye, and the insets sketch the quench scheme used. The position where the green line intersects the real-time axis, as indicated by black arrow, corresponds to the critical time $t_c$ where dynamical singularity occurs.}
	\label{fig7}
\end{figure}

The celebrated LY zeros provide a powerful framework to understand thermal phase transitions~\cite{yang1952,lee1952,fisher1965} 
and also DQPT~\cite{heyl2013}. In the latter case, one considers the boundary partition function $Z(z) = \langle \psi_i | e^{-zH_f} | \psi_i\rangle$
\cite{leclair1995,heyl2013} with $z = \tau + it \in \mathbb{C}$, and DQPT occurs when the LY-zero lines of $Z(z)$ cross the imaginary-$z$ 
(i.e., real-time) axis~\cite{heyl2013}. Here, we calculate $Z(z)$ for quantum quenches both across and not across the QSC boundary. 
Using the time-dependent variational principle algorithm~\cite{haegeman2011,haegeman2016} (more details are given in appendix), we study a chain of length $L = 32$ 
with $g_i = 0$ (i.e., $|\psi_i\rangle = | \uparrow, \uparrow, \dots, \uparrow \rangle$) at fixed $h = 0.01$. In Fig.~\ref{fig7}(a), for $g_f=0.4 < 
g_{\text{x}}$, LY zeros (green dots) do not cross the real-time axis; while in Fig.~\ref{fig7}(b), for $g_f=0.65 > g_{\text{x}}$, the LY zeros, which coalesce into lines in the thermodynamic limit, intersecting the real-time axis. Moreover, our results in Figs.~\ref{fig4} and \ref{fig7} indicate that no matter the quantum quench is from left to right or vise versa, as long as it traverses the QSC crossover line, dynamical singularities appear. Thus, while being a crossover line in equilibrium properties, it serves as a ``phase boundary" in quantum dynamics.

Originally, LY zeros exist in a hypothetical complex parameter space and are not physically accessible in classical supercritical 
regime above $T_c$~\cite{ouyang2024}. 
Here, however, the real-time axis in the complex $Z(z)$ intersects the LY-zero line [see Fig.~\ref{fig7}(b)],
providing a natural explanation for the origin of the dynamical singularity.

\begin{figure*}
\centering
\includegraphics[width=.98\textwidth]{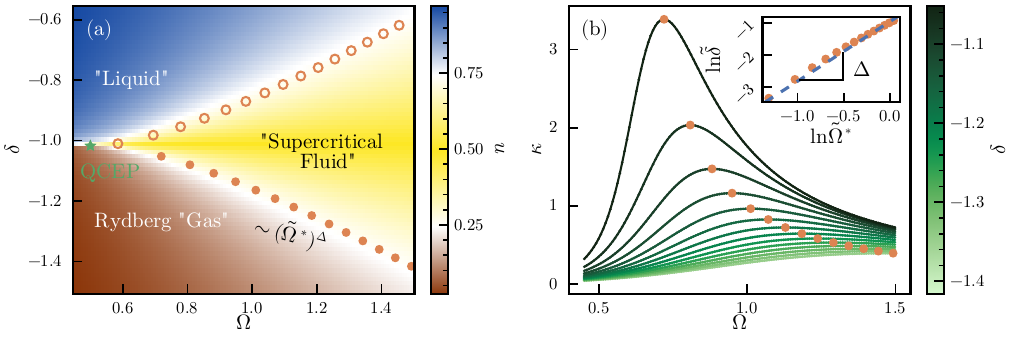}
\caption{\textbf{QSC crossover in Rydberg atom array.} (a) Rydberg excitation density $n$ as a function of laser detuning $\delta$ and Rabi frequency $\Omega$. Orange dots and circles are peaks of compressibility $\kappa=\partial n/\partial \delta$ at each fixed $\delta$, and the green star indicates the location of QCEP. (b) The peaks of $\kappa$ determine the QSC crossover line, and the inset shows the scaling law $\delta \sim (\tilde{\Omega}^*)^\Delta$ with $\tilde{\Omega}^* = \Omega - \Omega_c^*$.}
\label{fig8}
\end{figure*}

\section{Simulations of QSC crossover with Rydberg atom array}
We now introduce a proposal for experimental realization of quantum supercriticality, i.e., using the Rydberg atom array system~\cite{schauss2015,Browaeys2016,bernien2017,Guardado-Sanchez2018,omran2019,keesling2019,bluvstein2021,semeghini2021,ebadi2021,scholl2021}. Consider its standard Hamiltonian~\cite{schauss2018,browaeys2020,morgado2021,wu2021},
\begin{equation}
	H_{\text{Ryd}} = \frac{\Omega}{2}\sum_i \sigma_i^x - \delta \sum_i n_i + \sum_{i<j} V_{ij} n_i n_j,
	\label{rydberg}
\end{equation}
where $\Omega$ is the Rabi frequency that characterizes the coupling between ground state $|g_i\rangle$ and Rydberg state $|r_i\rangle$. $n_i$ is the corresponding Rydberg state number, $\delta$ is the laser detuning, and $V_{ij}=\frac{C_6}{(a|i-j|)^6}$ is the van der Waals interaction. $|C_6/a^6| \equiv 1$ is set as the energy unit, with $C_6$ the coupling coefficient and $a$ the nearest-neighbor distance. Here we consider $C_6<0$, which corresponds to the FM Ising system.

We perform DMRG~\cite{white1992,schollwoeck2011,fishman2022} calculations on a moderate size $L=64$, which is currently accessible for Rydberg-atom simulators~\cite{Guardado-Sanchez2018,ebadi2021,bluvstein2021,scholl2021} (more details on the method are given in appendix). Figure~\ref{fig8}(a) shows the supercritical phase diagram 
determined from the mean Rydberg excitation density $n = \sum_i \langle n_i \rangle/L$. The Rydberg supercritical ``fluid'' is enclosed by 
the two crossover lines, obtained from peaks of compressibility $\kappa \equiv \partial_{\delta} n $ at various fixed $\delta$ [see Fig.~\ref{fig8}(b)]. The crossover line, distinguishing the Rydberg atom ``gas'' and ``liquid'' from the supercritical ``fluid'' extends relatively far from the QCEP and follows the QSC scaling law $\delta \sim (\tilde{\Omega}^*)^\Delta$ [inset of Fig.~\ref{fig8}(b)]. Note since the long-range couplings decay algebraically as $1/r^\alpha$ with an exponent $\alpha = 6 > 3$, the QCEP also belongs to the (1+1)D Ising universality class~\cite{fisher1972,knap2013,fey2016,defenu2017}.
Moreover, we note that, by exploiting the $\mathbb{Z}_2$ symmetry of the QSC crossover lines, one can focus on the lower branch --- i.e., the solid dots in Fig.~\ref{fig8}(a) --- to study the QSC scaling law in experiments. With such a design, the majority of atoms remain in the ground state, effectively avoiding the avalanche dephasing effect induced by the black-body radiation.

\section{Discussions}
The CEP and associated supercriticality are fundamental concepts in studying phase transitions and critical phenomena. However, it has rarely been explored in the context of quantum phase transitions --- particularly QCEPs at zero temperature, despite their frequent occurrence in quantum many-body systems and materials. Here we reveal the existence of QSC crossover lines obeying the scaling law $h \sim \tilde{g}^\Delta$, through studying a prototypical quantum Ising model with QCEP. Despite smooth variation in equilibrium and entanglement properties, we identify singularities in quantum quench dynamics across the QSC crossover line. Significantly, we identify a new critical exponent (1/2 instead of 1) emerging from dynamical singularities associated with the QSC crossover --- a discovery that challenges existing understanding of dynamical criticality and demands new universal descriptions. We note that previous studies have also discussed DQPTs in the MFQIM~\cite{heyl2013,karrasch2013,denicola2021}: for $h=0$, linear cusp is found when quench across $g_c$; for $h\neq 0$, linear cusp is also observed with a suitable quench scheme. Thus, it is argued that DQPTs do not necessarily have a one-to-one correspondence with underlying equilibrium quantum phase transitions~\cite{vajna2014,andraschko2014}. Here, nevertheless, we demonstrate their intimate connection by revealing that the singular cusps occur precisely at the QSC crossover line associated with the QCEP. Lastly, we note that the QSC crossover lines determined through various methods --- both in and out of equilibrium --- do not coincide in absolute value. Nonetheless, all these lines follow the same QSC scaling law, which constitutes one of the main findings of this work.

Beyond the equilibrium properties of QSC crossover accessible by the Rydberg atom array as mentioned above, the non-equilibrium aspects of QSC crossover, such as the singular cusp in LRF, are also expected to be observed in various experimental setups.
For example, DQPT has been directly detected in a trapped ion experiment~\cite{jurcevic2017}. Since very recently this system has been used to study dynamics of 1D MFQIM~\cite{luo2025}, we anticipate that quench dynamics and LRF for MFQIM can also be readily accessed. Further, phenomena of DQPT have been studied in nuclear magnetic resonance simulator~\cite{nie2020}, optical lattice 
systems~\cite{flaschner2018}, topological nanomechanical systems~\cite{tian2019}, nitrogen-vacancy centers in diamond~\cite{yang2019,wang2019a}, quantum walks~\cite{wang2019b,xu2020}, and superconducting qubits~\cite{guo2019}. This demonstrates the broad applicability of our quantum supercriticality findings and motivates further exploration of supercritical dynamic singularities in quantum systems.

There are several promising future directions for quantum supercriticality studies. One is to explore physical implications of QSC crossovers at finite temperature. This is directly related to quantum materials like 1D Ising chain compound CoNb$_2$O$_6$~\cite{coldea2010, kinross2014, wu2014, liang2015, zou2021, xu2022, lv2024} and 3D Ising magnets LiREF$_4$ (with RE the rare earth elements)~\cite{Xie2021Giant, Wendl2022Mesoscale, Liu2023Ultralow}. In these systems, an external magnetic field can be used to tune the system across the QSC crossover, and the QSC scaling law can be verified by measuring magnetic susceptibility or specific heat~\cite{lv2024}. Also, unique quantum universality classes, such as the superfluid-insulator transition~\cite{continentino2017} and deconfined QCEP~\cite{senthil2004}, provide novel perspectives and opportunities for exploring supercritical phenomena.

%
%

\begin{acknowledgments}
J.W. and W.L. are indebted to Tao Shi, Ning Xi, Guoliang Wu, Lei-Yi-Nan Liu, Jian Cui and Haiyuan Zou for stimulating discussions.
This work was supported by the National Key Projects for Research and Development of China with Grant No. 2024YFA1409200, the National Natural Science Foundation of China (Grant Nos.~12222412 and 12047503), the Fundamental Research Funds for the Central Universities, and the Strategic Priority Research Program of Chinese Academy of Sciences through Grant No. XDB1270100. We thank the HPC-ITP for the technical support and generous allocation of CPU time.
\end{acknowledgments}





\appendix

\begin{figure*}[h] 
	\centering
	\includegraphics[width=.98\textwidth]{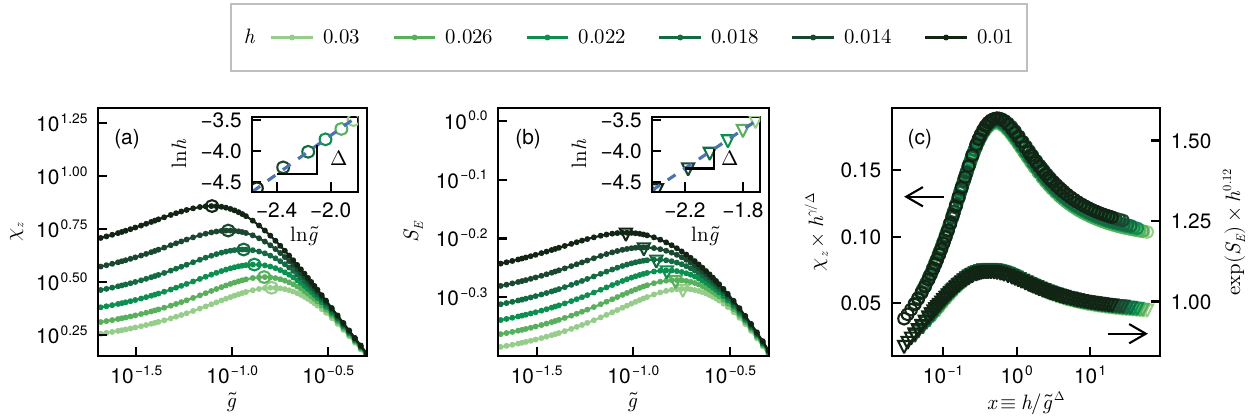} 
	\caption{\textbf{Determination of QSC crossovers of 2D MFQIM.} magnetic susceptibility $\chi_z$ (a) and entanglement entropy $S_E$ (b) used to determine the QSC crossover for the 2D MFQIM. In the insets, the hollow symbols represent the corresponding peaks, demonstrating a power-law scaling $h \sim \tilde{g}^{\Delta}$, as indicated by the blue dashed line, with the gap exponent $\Delta = \beta + \gamma \approx 1.563$. (c) data collapse for these quantities. The peaks collapse to a single point at $x = x_0$, which is the maximum of the corresponding universal scaling function.}
	\label{fig:2d} 
\end{figure*}

\section{Methods}
In this work, we use four types of Matrix Product State (MPS) tensor network methods to study various aspect of quantum supercriticality.
Note that they work in distinct scenarios, and together provide a comprehensive understanding of the QSC phenomena.

The first, in order of appearance, is the variational uniform matrix product state (VUMPS) algorithm~\cite{zaunerStauber2018}. It is an algorithm designed to efficiently find the ground state of infinite 1D and 2D cylinder (with finite width) systems. In practice, we use fixed bond dimension $D=512$ for 1D and $D=1024$ for the 2D case. For the 1D case, the local MPS tensor is a single-site one. And for the 2D case, we consider a width-12 cylinder, hence the local MPS tensors contains 12 sites. The convergence behavior vs. bond dimension for the 2D case is given in the last section of appendix. We use the package \textsc{MPSKit.jl}~\cite{damme2024}.

The second method is the infinite time evolution block decimation (iTEBD) algorithm~\cite{vidal2007,orus2008}. It is a numerical technique for simulating the real-time dynamics of quantum many-body systems represented by MPS. In this work, we use iTEBD to calculate the Loschmidt rate function $r(t)$ for quantum quenches across the QSC crossover line. The time step is chosen to be 1E-4 to capture the singular cusp in LRF, and we find a maximal bond dimension below $D=42$ is sufficient. We also use a bilayer trick, namely only evolve the system to $t/2$, and take the complex conjugate of the resulting MPS, then take the overlap with the original MPS. In this way, we obtain the Loschmidt amplitude at $t$.

The third method is the time-dependent variational principle (TDVP) algorithm~\cite{haegeman2011,haegeman2016}. It is a numerical technique for computing the time evolution of quantum states represented by MPS. This method is particularly useful for studying dynamical properties of quantum systems, such as DQPT. In this work, we use TDVP to calculate the LY zeros of the boundary partition function $Z(z)$ for quantum quenches across and not across the QSC crossover line. The system size used is $L=32$ and the bond dimension is set to $D=20$. We use the package \textsc{FiniteMPS.jl}~\cite{li2025}.

The last method is the density matrix renormalization group (DMRG) algorithm~\cite{white1992,schollwoeck2011}. It is suitable for finding the ground state of finite-size quantum many-body systems in 1D. In this work, we use DMRG to study the Rydberg atom array system given by Eq.~\eqref{rydberg}. The system size is set to $L=64$, and we use a bond dimension of $D=128$. The interaction $V_{ij}$ is truncated up to the fifth neighboring sites, which is justified as $V_{ij}\sim 1/r^6$ decays rapidly. We use the 
package \textsc{ITensors.jl}~\cite{fishman2022}.

\begin{figure*}[h] 
	\centering
	\includegraphics[width=.98\textwidth]{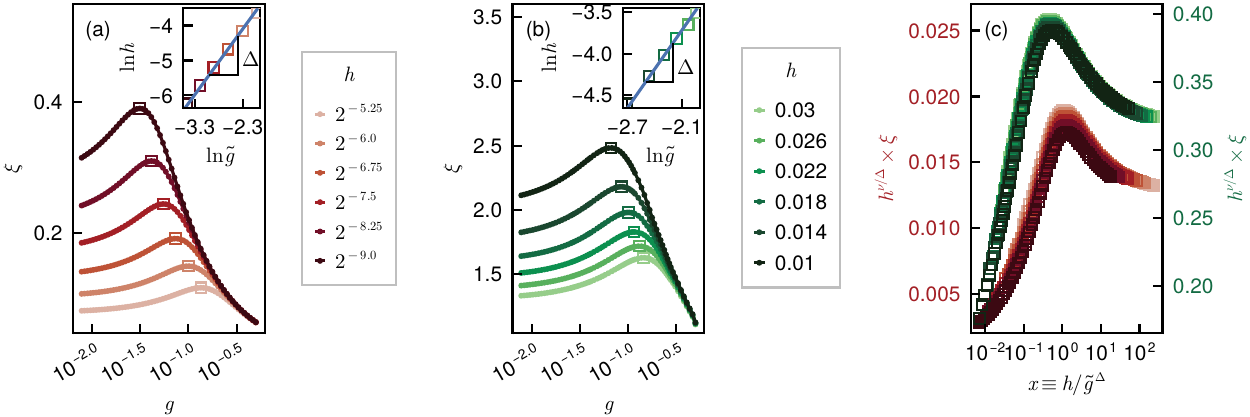} 
	\caption{\textbf{Determination of QSC crossovers of 1D and 2D MFQIMs.} QSC crossovers determined by using correlation length $\xi$ for 1D (a) and 2D (b) MFQIMs. (c) shows their data collapse respectively.}
	\label{fig:sm_fig1} 
\end{figure*}

\section{Further results on the 1D and 2D MFQIM: equilibrium aspect}
We present magnetic susceptibility $\chi_z$ and entanglement entropy $S_E$ for the 2D MFQIM in Fig.~\ref{fig:2d}. Note the QCEP now is located at $g_c \approx 1.52$~\cite{blote2002} with $h=0$. We employ the variational uniform matrix 
product state algorithm~\cite{zaunerStauber2018} to simulate 2D infinite cylinders with width $W=12$. For both quantities, all curves initially exhibit a divergence behavior before eventually becoming suppressed. Collecting all these peaks in the inset, it demonstrates that this line follows a 
power-law scaling $h \sim \tilde{g}^{\Delta}$ with $\Delta =\beta+\gamma \approx 1.563$. Although the width $W=12$ is still limited, a finite $h$ introduces a cutoff in correlation length, alleviating the finite-size effect and rendering the robust 2D QSC scaling. We further understand this equilibrium QSC scaling law from universal scaling functions similar to the 1D case. The data collapse results are given in Fig.~\ref{fig:2d}(c), the maximal point of these universal scaling functions corresponds precisely to the QSC crossover line. Note here $\exp(S_E) \sim h^{-y} 
\phi_{S_E}(x)$ is conjectured for the 2D case, and we numerically find that excellent collapse is achieved with fitted $y=0.12$. A renormalization group analysis is needed in the future to confirm this scaling form.

We further present additional results supporting the QSC scaling shown in Fig.~\ref{fig:sm_fig1}, namely the correlation length data. The 1D case and 2D case is presented in Fig.~\ref{fig:sm_fig1}(a) and (b), respectively. The inset illustrates that the maximum locations of $\tilde{g}$ adhere to the QSC scaling law $h \sim \tilde{g}^\Delta$. To understand this, we note that the singular part of correlation length $\xi$ close to the QCEP can be described by a generalized homogeneous function~\cite{fisher1974,pelissetto2002,hankey1972,kirkpatrick2015}, $\xi \sim \tilde{g}^{-\nu}\phi_{\xi}(x)$, where $x\equiv h/\tilde{g}^{\Delta}$. This relation leads to the data collapse of $\xi$ shown in Fig.~\ref{fig:sm_fig1}(c). Thus the peak of $\xi$ corresponds to a constant $x = x_0$ such that $\phi_{\xi}'(x_0)=0$, which leads to QSC crossovers satisfying $h/x_0 = \tilde{g}^{\Delta}$.

\begin{figure*}[h] 
	\centering
	\includegraphics[width=0.98\textwidth]{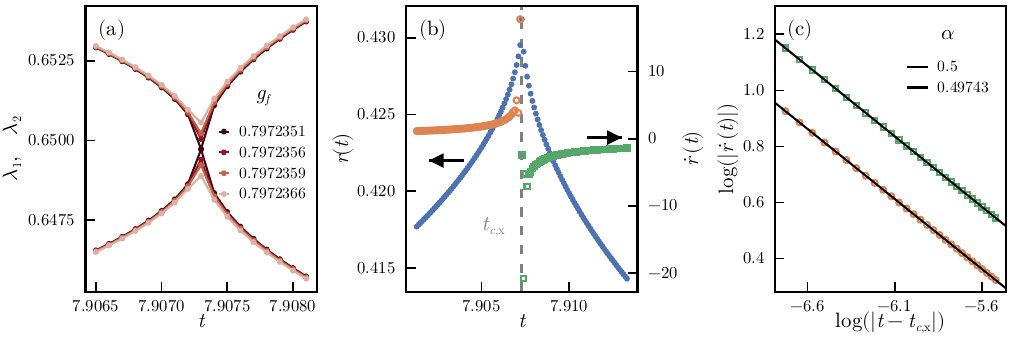} 

	\caption{\textbf{Scaling behavior of DQPT near the critical time $t_{c,\rm{x}}$.} (a) Two dominant eigenvalues $\lambda_1, \lambda_2$ of the mixed transfer matrix around the critical time for different $g_f$. (b) The Loschmidt rate function $r(t)$, and its time derivative $\dot{r}(t)$ for quantum quench to $g_f = g_{\text{x}} \simeq 0.7972351$. (c) The log-log plot of $\dot{r}(t)$ near $t_{c,{\rm x}}$, from which we determine the power $\alpha \simeq 1/2$ as defined in Eq.~\eqref{eq:r_scaling}. Other parameters used: $h=0.08$ and $D=20$.}
	\label{fig:fig5} 
\end{figure*}

\section{The Loschmidt-rate critical exponent at $t_{c,\rm{x}}$}
As shown in the left top inset of Fig.~\ref{fig4}(a), the time derivative of Loschmidt rate behaves differently for $g_f < g_{\text{x}}$ and $g_f = g_{\text{x}}$. For the former case, the time derivative $\dot{r}(t)$ exhibits a discontinuous jump at $t_c$, i.e., it converges to two different constants from both sides, which leads to a linear cusp. On the other hand, for the latter case $g_f = g_{\text{x}}$, $\dot{r}(t)$ shows instead a divergent peak, i.e., $\dot{r}(t)|_{t \to t_{c, {\rm x}}} = \pm \infty$ from two sides of $t_{c, {\rm x}}$, which is a 1/2 cusp.

To carefully characterize such dynamical singularities, we expand the Loschmidt rate function $r(t)$ near the critical time $t_{c}$ as
\begin{equation}
    r(t) \simeq F^\pm |t - t_c|^{\alpha_\pm} + r(t_c).
    \label{eq:r_scaling}
\end{equation}
For $g_f < g_{\rm x}$, we have linear scaling behavior with $\alpha_\pm = \alpha = 1$.
This can be understood as a level-crossing between two dominant eigenvalues, $\lambda_1$ and $\lambda_2$, 
of the mixed transfer matrix~\cite{karrasch2013,andraschko2014,zauner2017}. The transfer matrix can be obtained 
as $\mathcal{T}(t) = \sum_s \bar{A}^s(0) \otimes A^s(t)$, for two local tensors $A^s(0)$ and $A^s(t)$ of the uniform 
matrix product states, at time zero and $t$, respectively.

For the singular cusp $g_f = g_{\rm x}$, as illustrated in Fig.~\ref{fig:fig5}(a), the two dominant eigenvalues 
merely ``kiss" each other, undergoing an avoided level crossing. In Fig.~\ref{fig:fig5}(b), we present the 
Loschmidt rate function and its time derivative $\dot{r}(t)$ in the vicinity of $t_{c, {\rm x}}$, where a divergent cusp is evident. Notably, both the positive and negative branches of $\dot{r}(t)$ exhibit a power-law behavior with the same exponent $-1/2$, which indicates that $\alpha \simeq 1/2$ [see Fig.~\ref{fig:fig5}(c)]. This distinctive feature of a singular cusp with critical exponent $1/2$ was previously derived from a renormalization group analysis in Ref.~\cite{wu2020}. Our findings indicate that such singular cusps can emerge systematically in quenches to the QSC crossovers, as also verified for quantum Potts model discussed in the next section.

\begin{figure*}[h]
	\centering
	\includegraphics[width=.98\textwidth]{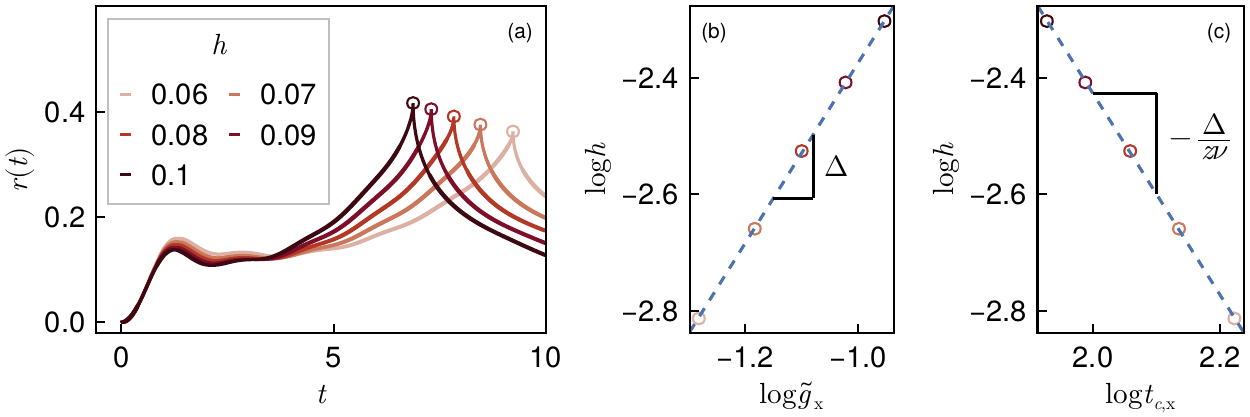}
	\caption{\textbf{Dynamical singularity of QSC crossover in the 1D 3-state MFQPM.} (a) LRF with a singular cusp at $t_{c,\text{x}}$ for various $h$ and $g_f = g_{\text{x}}(h)$. (b) Log-log plot of $h$ vs. $\tilde{g}_{\text{x}}$, showing the QSC scaling law $h \sim \tilde{g}_{\text{x}}^\Delta$. (c) Log-log plot of $h$ vs. $t_{c,\text{x}}$, illustrating a scaling relation $h \sim (1/t_{c,\text{x}})^{\Delta/z\nu}$. Note that for 3-state QPM, $\Delta = 14/9$, $z=1$ and $\nu = 5/6$.}
	\label{fig6}
\end{figure*}

\section{Universal dynamical singularity in quantum Potts model}
As a validation of the universal aspect of this quench dynamics beyond MFQIM, we consider the mixed-field quantum Potts model (MFQPM) with Hamiltonian,
\begin{equation}
	H = - J \sum_i \sum_{\mu=1}^q P_i^\mu P_{i+1}^\mu
	- g \sum_i P_i - h\sum_i \sum_{\mu=1}^q \frac{1}{\mu} P_i^\mu.
    \label{potts}
\end{equation}
Here, the traceless operator $P_i^\mu = |\mu_i\rangle\langle\mu_i| - 1/q$ represents the $\mu$-th component of $Z_q$ spin at site $i$, thus the first term favors an ``FM'' ground state by breaking the $Z_3$ or $Z_4$ symmetry for the $q=3$ and $q=4$ 
cases, respectively. The $g$ term represents a transverse field, with $P_i = |\lambda_i \rangle \langle \lambda_i | - 1/q$ and $|\lambda_i \rangle = \sum_{\mu_i} |\mu_i \rangle /\sqrt{q}$, which drives the system to a paramagnetic (PM) phase. The $h$ term represents a 
longitudinal field, which explicitly breaks the permutation symmetry $S_q$ and acting on all components with a biased
weight $1/\mu$. Alternatively, one can also simply add $-h\sum_i P^\mu_i$ to select a specific $\mu$ component of the $Z_q$ spin. Similar to the MFQIM, a QCEP between the FM and PM phases exists at $g_c = 1$ and $h = 0$. There are also QSC crossovers for $h \neq 0$, and the equilibrium aspects are thoroughly studied in the next section. Here we study the quantum quench dynamics, with exactly the same setting as the MFQIM. Namely, we prepare the state as the ground state of the Hamiltonian with $g_i = \rightarrow \infty$, i.e., $| \psi_i \rangle = \otimes_j | \lambda_j \rangle$, and quench the Hamiltonian to various $g_f$ with a fixed small $h$. Remarkably, we again find that if quench across the QSC crossover line, there is linear cusps, and if quench precisely onto the QSC crossover line,
there is 1/2 singular cusps. These results are summarized in Fig.~\ref{fig6}(a) and (b) for the 3-state MFQPM. Additionally, we also find that the critical time obeys the same scaling relation, similarly as the MFQIM, as shown in Fig.~\ref{fig6}(c). These results demonstrate that the dynamical singularity is indeed a universal property of QSC crossover, as discussed in Sec.~\ref{seclrf}.

\begin{figure*}[h] 
	\centering
	\includegraphics[width=.98\textwidth]{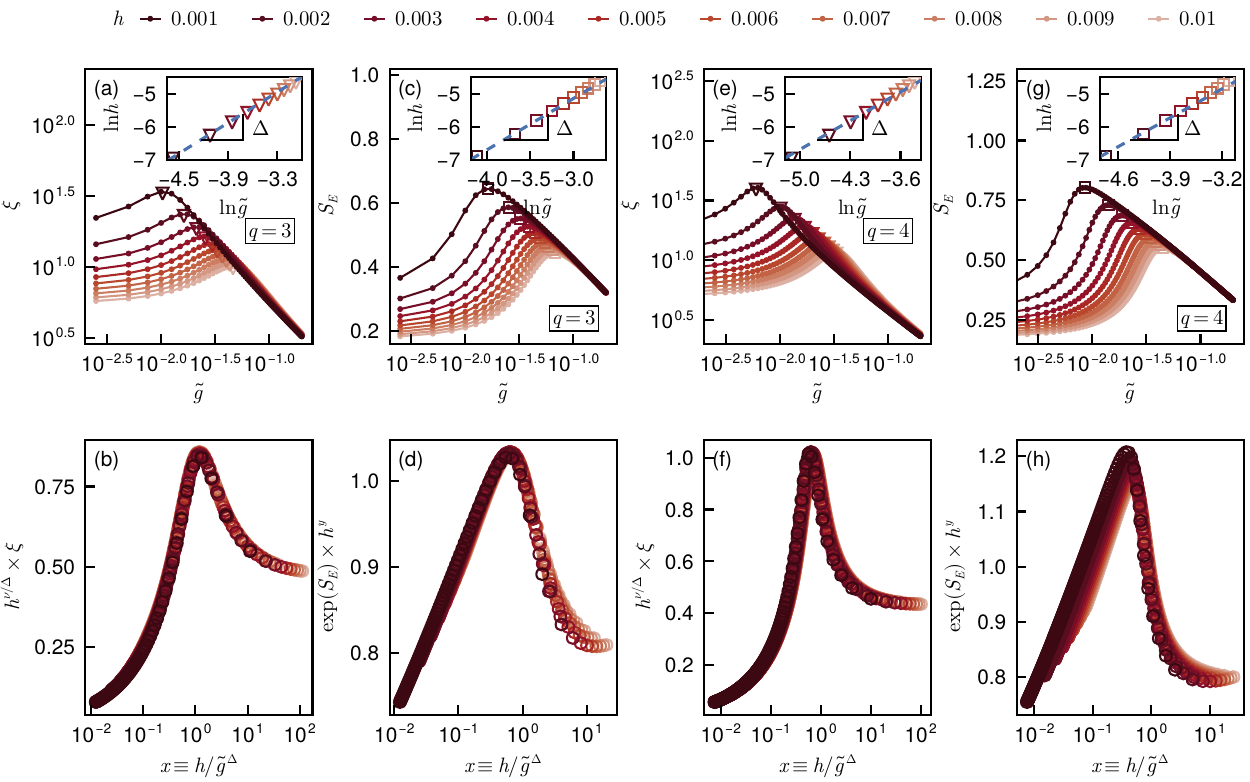} 

	\caption{\textbf{Determination of QSC crossovers of 1D MFQPM.} The QSC crossovers of QPM is shown for $q=3$ in (a) and (c), and for $q=4$ in (e) and (g), near the QCEP, from the correlation length $\xi$ and entanglement entropy $S_E$, respectively. The hollow markers indicate peaks of each curve, which exhibit the power-law scaling $h \sim \tilde{g}^{\Delta}$, with $\tilde{g} \equiv g-g_c$, as illustrated in corresponding insets. The second line (b), (d), (f) and (h) show the data collapse of $\xi$ and $S_E$, respectively.}
	\label{fig:sm_fig2} 
\end{figure*}

\section{QSC crossovers in the $q$-state MFQPM: equilibrium aspect}
Here we also consider the equilibrium aspect of QSC crossover in the $q$-state (for $q$=3 and 4) MFQPM, whose Hamiltonian is given in Eq.~\eqref{potts}. We employ the VUMPS algorithm \cite{zaunerStauber2018} to calculate the ground-state properties of 1D infinite chain. From the behaviors of the correlation length $\xi$ [Fig.~\ref{fig:sm_fig2}(a) and (e)] and the entanglement entropy 
$S_E$ [Fig.~\ref{fig:sm_fig2}(c) and (g)], we reveal the QSC crossovers with universal scalings for both MFQPMs (shown in the respective insets). Note that the corresponding gap exponents are $\Delta = 14/9$ and $\Delta = 5/4$, for $q = 3$ and $q = 4$, respectively. We further comprehend these results by performing data collapse, as shonw in Fig.~\ref{fig:sm_fig2}(b), (d), (f) and (h).

\begin{figure*}[h] 
	\centering
	\includegraphics[width=0.7\textwidth]{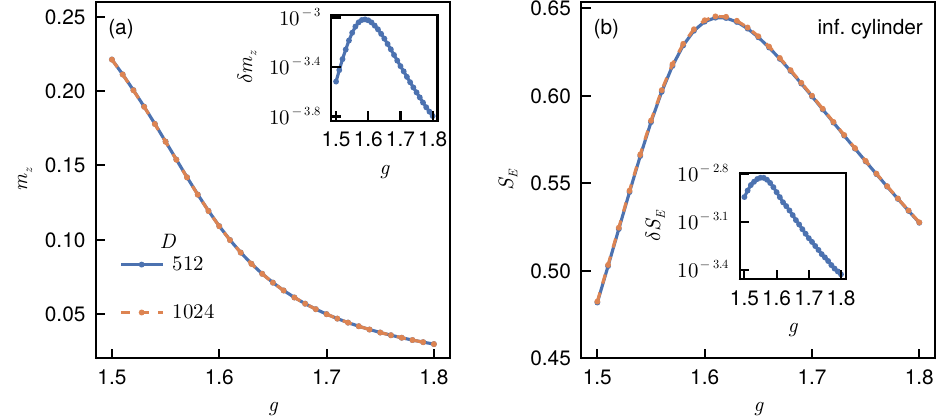} 

	\caption{\textbf{Convergence check for 2D calculations.} (a) Longitudinal magnetization $m_z$ and (b) entanglement entropy $S_E$ vs. transverse field $g$, for two bond dimensions $D=512$ and $1024$, for the 2D MFQIM on an infinite-long cylinder with width $W=12$, and under a fixed longitudinal field $h/J=0.01$. The insets show the relative difference between two bond dimensions, e.g., $\delta m_z \equiv |[m_z(D=1024) - m_z(D=512)] / m_z(D=1024)|$.}
	\label{fig:sm_fig3} 
\end{figure*}

\section{Convergence on results for the two-dimensional quantum Ising model}
In Fig.~\ref{fig:sm_fig3}, we present the calculated longitudinal magnetization $m_z$ and entanglement entropy 
$S_E$ vs. transverse field $g$ at a fixed longitudinal field of $h=0.01$. The simulations were carried out utilizing 
the VUMPS algorithm~\cite{zaunerStauber2018} for a system on an infinitely 
long cylinder with a finite width $W=12$. To assess numerical convergence, we compared results obtained using 
two different bond dimensions, $D=512$ and $D=1024$, and the insets illustrate a very good convergence.

Regarding other 2D cylinder results involved in this study, their respective longitudinal fields are greater than the 
value showcased in Fig.~\ref{fig:sm_fig3}. Owing to their greater distance from the QCEP, achieving convergence 
is expected to be more straightforward. Therefore, Fig.~\ref{fig:sm_fig3} demonstrates that the bond dimension $D=1024$ is appropriate to ensure a very good convergence and high accuracy of the calculations. In conclusion, all data for 
2D studies used in this work sticks to this bond dimension.

%

\end{document}